# Статистическая сумма для вибронной модели в неаддитивном случае (наноразмерные объекты)

Рассматривается задача построения статистической суммы для неаддитивного случая (для наноразмерного объекта) в представлении нанотермодинамики Хилла. Вынужденный отказ от постулата аддитивности приводит к задаче построения обобщённой статистической суммы на основе комбинаторно-статистического усреднения.

Энергетика объекта рассматривается на основе вибронной модели (подразделы «Вибронная модель» и «Структура матрицы гамильтониана») с использованием авторского способа учёта размерной зависимости количеств структурных единиц и элементов (подраздел «Размерная зависимость»). Далее вводится комбинаторно-статистическое усреднение матричных элементов гамильтониана объекта (подразделы «Комбинаторная статистика в матрице гамильтониана», «Комбинаторное усреднение матричных элементов»). Результатом усреднения являются выражения для энергий индивидуальных квантовых состояний объекта и его структурных элементов, а также выражение для статистической суммы объекта, допускающей факторизацию (подраздел «Решение»).

## Проблема

Одной из термодинамических методологий, приложимых к наноразмерным объектам, является нанотермодинамика Хилла [1]. Особого рода ансамбль, вводимый в этой методологии, будем здесь называть *хилловским коллективом*. Поскольку хилловский коллектив макроскопичен и по определению состоит из невзаимодействующих между собой *малых (хилловских) объектов*, для него справедливо основное соотношение статистической термодинамики между одним из термодинамических потенциалов и логарифмом статистической суммы [1]:

$$\Phi = -k_B T \ln \Xi_K = -k_B T \ln \Xi^{\mathbf{K}},$$

где $\Phi$ — потенциал; $\Xi_K$ — статистическая сумма для коллектива в целом; $\Xi$ — статистическая сумма для отдельного объекта; $\mathbf{K}$ — количество объектов в коллективе.

При этом статистическая сумма в методологии Хилла используется в обобщённом (по Хиллу) виде, то есть, уточнена добавлением множителей—статистических весов, зависящих от внутренних параметров. Так, например, обобщённая каноническая статистическая сумма в условиях $NVT$ [1, ч.1, §1-3, 1-4; 1, ч.2]:

$$\Xi_{NVT} = \sum_q \exp[-\beta \mathcal{E}_q(N,V)],$$

где $\mathcal{E}_q(N,V)$ — энергия квантового уровня $q$, формально зависящая от экстенсивных переменных $N$ и $V$; $N$ — количество структурных единиц (атомов, ионов, молекул) в объекте; $V$ — объём объекта; $\beta=(k_B T)^{-1}$ — температурный параметр.

Обобщённая статистическая сумма для хилловского объекта в условиях $NPT$ [1]:

$$\Xi_{NPT}=\sum_q \sum_\nu \exp(-\beta \mathcal{E}_q)\exp(-\beta p \mathrm{v}_\nu)=\sum_\nu \exp(-\beta p \mathrm{v}_\nu)Z, \qquad (1)$$

где $q$ — индекс энергетических состояний объекта; $\mathcal{E}_q$ — энергия состояния с индексом $q$; $\nu$ — индекс различных величин объёма объекта; $\mathrm{v}_\nu$ — объём объекта, обозначенный индексом $\nu$; $Z$ — каноническая статистическая сумма для целого объекта.

Далее — в традиционном подходе — полная статсумма (факторизуемая) разлагается на элементарные множители [2]:

$$\Xi_{\text{факт}}=\sum_\nu \exp(-\beta p \mathrm{v}_\nu)Z=\sum_\nu \exp(-\beta p \mathrm{v}_\nu)z^L, \qquad (2)$$

где $z$ — каноническая статистическая сумма для одного элемента (например, локальной моды); $L$ — количество элементов; .

Однако *предположение об аддитивности* [2], разрешающее такое разложение, в наноразмере оказывается некорректным, так что выражение (2) использовать нельзя, по крайней мере, без существенных дополнений.

## Вибронная модель

Одной из методологий, пригодных для изучения энергетических характеристик объектов в наноразмерном масштабе, является *вибронная модель* (*вибронный метод*, в современной англоязычной литературе часто просто *algebraic method*).

Основные результаты методологии, важные для решаемой задачи, это форма оператора Гамильтона для объекта (операторное уравнение) и соответствующее матричное уравнение.

Итоговый вид гамильтониана в т. наз. одномерном варианте (с использованием только одномерных квантовых вибраторов) для физического объекта, состоящего из $L$ взаимодействующих локальных мод (квантовых вибраторов Морза) [3]:

$$\mathcal{H}=\mathcal{E}_0+\sum_i^L A_i \mathcal{C}^{(2)}_{AO_i(2)}+\sum_i^L \sum_{j>i}^L A_{ij}\mathcal{C}^{(2)}_{AO_{ij}(2)}+\sum_i^L \sum_{j>i}^L \lambda_{ij}\mathcal{M}_{ij}, \qquad (3)$$

где $\mathcal{E}_0$ — нулевая энергия объекта, т. е., сумма нулевых уровней отдельных мод; $\mathcal{C}$ — *операторы Казимира* — инвариантные операторы алгебр Ли AO(2); $\mathcal{M}$ — *операторы Майораны* — инвариантные операторы алгебр AU(2); $A_i$, $A_{ij}$, $\lambda_{ij}$ — параметры со смыслом весовых множителей; верхние индексы вида (2) обозначают вторую степень оператора [3].

Слагаемые вида $A_i \mathcal{C}^{(2)}_{AO_i(2)}$ фактически есть собственные значения энергий отдельных мод (вибраторов Морзе) при отсутствии взаимодействий. Параметры $A_i$ есть соответствующие силы потенциала $A$. Слагаемые вида $A_{ij} \mathcal{C}^{(2)}_{AO_{ij}(2)}$ и $\lambda_{ij} \mathcal{M}_{ij}$ выражают ангармонические сцепления (взаимодействия) между различными модами. Параметры $A_{ij}$ и $\lambda_{ij}$ описывают силу соответствующей взаимосвязи (сцепления). При существенности величины силы взаимосвязи по сравнению с параметром ангармоничности единичной моды энергетический спектр системы приближается к нормально-модовому [4]. При $\lambda_{ij}=0$ получается предел вполне локальных мод. Но и в этом случае в описании системы присутствует некоторая внутренняя корреляция, выраженная через слагаемые $A_{ij} \mathcal{C}^{(2)}_{AO_{ij}(2)}$ [3].

Одномерная потенциальная функция Морза в *лиевско-алгебраическом представлении* имеет вид [3]:

$$\mathcal{E}(m) = \mathcal{E}_0 + A m^2; \quad \mathcal{E}(\upsilon) = \mathcal{E}_0 + A(N-2\upsilon)^2 = \mathcal{E}'_0 - 4A(\Upsilon \upsilon - \upsilon^2), \qquad (4)$$

где $A$ — *сила потенциала*, параметр алгебраического представления; $\Upsilon$ — дополнительное квантовое число; $m$ — «алгебраическое» квантовое число, вводимое как $m=\Upsilon, \Upsilon-2, ..., 1$ для нечётных $\Upsilon$ и $m=\Upsilon, \Upsilon-2, ..., 0$ для чётных $\Upsilon$; $\upsilon$ — преобразованное алгебраическое квантовое (вибронное) число, вводимое как $\upsilon=(\Upsilon-m)/2$; вибронное число имеет пределы $\upsilon=0 ... \lfloor \Upsilon/2 \rfloor = 0 ... \eta$, где $\lfloor x \rfloor$ — округление до наибольшего целого, не превышающего $x$.

Таким образом, собственные значения гамильтониана получаются как алгебраические формы от собственных значений соответствующих инвариантных операторов. Ещё одним важным свойством построенного таким образом оператора Гамильтона при отсутствии взаимодействия со внешней средой оказывается сохранение значения полиады или полного вибронного числа, т. е., суммы вибронных чисел во всех локальных модах.

Обычно в вибронной модели работают с приближением, в котором учитываются межмодовые взаимодействия лишь в парах ближайших соседей (в пределах первых координационных сфер каждой пары атомов, составляющих одну моду). В этом случае, если допустить, что все моды одинаковы по составу, операторное уравнение (3) соответствует матричному уравнению

$$\begin{aligned}
\mathbf{H} = \mathbf{E}_0 &+ A(\mathbf{C}_1 + \mathbf{C}_2 + ... + \mathbf{C}_L) + \\
&+ A'/2 \left( \sum_{j=1}^{B} \mathbf{C}_{1j} + \sum_{j=1}^{B} \mathbf{C}_{2j} + ... + \sum_{j=1}^{B} \mathbf{C}_{Lj} \right) + \\
&+ \lambda/2 \left( \sum_{j=1}^{B} \mathbf{M}_{1j} + \sum_{j=1}^{B} \mathbf{M}_{2j} + ... + \sum_{j=1}^{B} \mathbf{M}_{Lj} \right)
\end{aligned} \quad (5)$$

где $\mathbf{H}$, $\mathbf{E}_0$, $\mathbf{C}_i$, $\mathbf{C}_{ij}$, $\mathbf{M}_{ij}$ — матрицы соответствующих операторов; $A$, $A'$, $\lambda$ — соответствующие множители инвариантов, причём $A$ — параметр в алгебраическом представлении потенциальной функции Морза; одинарные индексы матриц инвариантов перечисляют моды в составе тела; парные индексы перечисляют моды, сцепленные с каждой из мод; $B$ — количество мод, сцепленных с данной, определяемое по структуре объекта; параметры $A'$, $\lambda$ делятся на 2, поскольку при таком суммировании каждое сцепление учитывается дважды.

### Размерная зависимость

Предыдущие выражения включают количества структурных единиц (атомов, ионов, молекул) и структурных элементов (взаимодействий, локальных мод) в составе объекта.

Один из возможных методов перечисления структурных единиц и элементов, составляющих объект, предлагается в [5, 6]. Используемые здесь результаты это:

1) Размерно-зависимое количество локальных мод $L(N)$ в объекте; зависимость от размера выражена явно зависимостью от числа структурных единиц $N$ и неявно — зависимостью от формы и структуры объекта.

2) Размерный множитель $R(N)$, показывающий размерную зависимость количества локальных мод $L(N)$, приходящихся на одну структурную единицу:

$$R(N) = \frac{L(N)}{N}. \quad (6)$$

Размерный множитель имеет макроскопический предел, который есть постоянная для данного типа упаковки:

$$\lim_{N \to \infty} R_{yn}(N) = \mathcal{R}_{yn}, \quad (7)$$

а именно, 14 (6+8) для простой кубической, 12 для объёмноцентрированной и гранецентрированной кубических решёток.

3) Размерно-зависимое количество сцепленных (обменивающихся энер-

гией) локальных мод $L_L(N)$. В приближении взаимодействия ближайших соседей:

$$L_L(N) \simeq N R^2(N)/2. \tag{8}$$

## Структура матрицы гамильтониана

В целом матрица оператора Гамильтона в вибронной модели имеет *блочно-диагональную* (*ступенчатую*) структуру, где каждый блок соответствует одной *полиаде* (суммарному вибронному числу) [3].

Приняв указанное упрощение общего вида модели, получаем, что каждый элемент матрицы **H** соответствует определённому распределению вибронных чисел $n_j$ в наборе локальных мод, составляющих объект:

$$\{n_j\},\ j=1 \ldots L(N), \tag{9}$$

и алгебраически складывается из следующих частей:

1) «Диагональное» слагаемое, определяемое суммой собственных значений инвариантов Казимира $\mathcal{C}_i$ алгебр $AO_j(2)$ [3]:

$$\varphi_{c,ag} = A \sum_{i=1}^{L} \langle \mathcal{C}_i \rangle, \tag{10}$$

$$\langle v_i | \mathcal{C}_i | v_i \rangle = -4(Nv_i - v_i^2), \tag{11}$$

где индексы $a$, $g$ есть координаты элементов локально-модового базиса в матрице **H**.

2) «Диагональное» слагаемое, определяемое суммой функций от парных сочетаний собственных значений инвариантов Казимира $\mathcal{C}_{ij}$ алгебр $AO_{1:L}(2)$ [3]:

$$\varphi_{cd,ag} = \frac{A'}{2} \sum_{i=1}^{L} \sum_{b} \langle \mathcal{C}_{ij} \rangle,$$

$$\langle v_i v_j | \mathcal{C}_{ij} | v_i v_j \rangle = 4 \left[ (v_i + v_j)^2 - 2\Upsilon(v_i + v_j) \right], \tag{12}$$

где $b$ — индекс, перечисляющий (извлекающий) номера (метки) $j$ мод, сцепленных с модой, имеющей метку $i$ (иначе говоря, индекс для $i$-го множества мощности $B$).

3) «Диагональное» слагаемое, определяемое суммой функций от парных сочетаний собственных значений инвариантов Майораны $\mathfrak{M}_{ij}$ алгебр $AU_j(2)$ [3]:

$$\varphi_{md,ag} = \frac{\lambda}{2} \sum_{i=1}^{L} \sum_{b} \langle \mathfrak{M}_{ij} \rangle,$$

$$\langle v_i v_j | \mathfrak{M}_{ij} | v_i v_j \rangle = \Upsilon(v_i + v_j) - 2 v_i v_j. \qquad (13)$$

4) «Внедиагональное» слагаемое, определяемое суммой функций от парных сочетаний собственных значений инвариантов Майораны $\mathfrak{M}_{ij}$ [3]:

$$\phi_{mm,ag,a'g'} = \frac{\lambda}{2} \sum_{i=1}^{L} \sum_{b} \langle \mathfrak{M}_{ij} \rangle, \qquad (14)$$

где

$$\langle v_i+1; v_j-1 | \mathfrak{M}_{ij} | v_i; v_j \rangle = -\left[(v_i+1)(\Upsilon-v_i)\cdot v_j(\Upsilon-v_j+1)\right]^{1/2},$$

$$\langle v_i-1; v_j+1 | \mathfrak{M}_{ij} | v_i; v_j \rangle = -\left[v_i(\Upsilon-v_i+1)\cdot(v_j+1)(\Upsilon-v_j)\right]^{1/2}, \qquad (15)$$

где индексы $a$, $g$, $a'$, $g'$ есть координаты пары элементов локально-модового базиса в матрице **H**; вообще говоря, как видно из (15),

$$\phi_{mm,ag,a'g'} \neq \phi_{mm,a'g',ag}.$$

Отдельные слагаемые в (14) имеют ненулевое значение, если разницы вибронных чисел в модах $i$, $j$ составляют $\pm 1$, т. е., если моды обмениваются ровно одним виброном Сумма, построенная по образцу (14), входит в каждый элемент $h_{ag}$ матрицы **H**, для которого существует хотя бы одна ненулевая величина (15), но только в пределах блока, соответствующего одной и той же полиаде $p$ (все ограничения, о которых говорится в этом абзаце, вытекают из способа определения майорановских инвариантов [3]).

5) «Диагональное» слагаемое, составленное величинами энергий нулевых уровней структурных элементов (мод). Далее не показывается.

Таким образом, в матрице **H** ненулевые элементы $h_{ag}$, $a \neq g$ строятся как суммы

$$h_{ag} = \sum_{k, a'g' \neq ag} \phi_{mm,ag,a'g'}, \qquad (16)$$

где $k$ — индекс, перечисляющий элементы базиса, имеющие координаты $a'g'$ в матрице **H**, и входящие в тот же полиадный блок $p$.

Элементы $h_{ag}$, $a=g$ в матрице **H** строятся как суммы

$$h_{ag}=\varphi_{c,ag}+\varphi_{cd,ag}+\varphi_{md,ag}. \qquad (17)$$

Величина $B$ в (5) определяется как зависимость от размерного множителя $R(N)$:

$$B(N)=R(N)-1, \qquad (18)$$

и также является размерно-зависимой.

## Комбинаторная статистика в матрице гамильтониана

Полиада $Q$ объекта в вибронной модели есть сумма вибронных чисел $q_m$ в отдельных модах (квантовых вибраторах): $Q=\sum_{m=1}^{L} q_m$, и её величина есть целое число, пробегающее значения $Q=0...L\eta$, где $\eta$ — предельное значение вибронного числа в одной моде. Каждый блок матрицы соответствует одному значению полиады $Q_j$, и каждый столбец и строка матрицы в пределах блока соответствуют одному элементу базиса гамильтониана, который формально, но взаимно однозначно соответствует одной из всех возможных *композиций* [7] числа $Q_j$, причём таких, в которых 1) количество частей равно некоторому значению $k$, и 2) размеры (величины) частей не превышают некоторого значения $l$, и 3) могут содержаться части нулевого размера; такой вид композиции называется также *композицией в прямоугольнике*, по виду её *диаграммы Феррера* [8] (далее в этой статье — *прямоугольная композиция*).

Согласно физическому смыслу задачи, количество частей зафиксировано как $L=L(N)$; размеры частей находятся в пределах $0...\eta$. Структура заполнения блока матрицы, соответствующего полиаде $Q_j$, однозначно соответствует структуре набора прямоугольных композиций $Q_j$, содержащего все возможные композиции такого рода.

Вообще комбинаторная статистика для композиций, по сравнению со статистикой для разбиений, разработана несколько хуже [9, 10, 11]; далее кратко изложены основные результаты, относящиеся к решаемой задаче.

Количество прямоугольных композиций. Количество прямоугольных композиций для числа $n$, т. е., зависимость $\kappa(n,k,l)$, выражается точно через *полиномиальный (расширенный биномиальный) множитель* [12, p.77], являющийся обобщением *биномиального множителя* [13], и обычно обозначаемый как

$$\binom{k,l+1}{n}. \tag{19}$$

Величина $\kappa(n,k,l)$ при зафиксированных ограничениях $k,l$ образует, при пробегании аргументом интервала $0\ldots n_{max}$, *унимодальную последовательность* [14, 8], т. е., всегда существует такое число $n_0$, $0\leq n_0\leq X$, что

$$\kappa(0)\leq\kappa(1)\leq\ldots\kappa(n_0-1)\leq\kappa(n_0)\geq\kappa(n_0+1)\geq\ldots\geq\kappa(n_{max}), \tag{20}$$

причём известно, что [14]

$$n_0=\lfloor k(l+1)/2 \rfloor, \tag{21}$$

где $\lfloor x \rfloor$ — наибольшее целое число, не превышающее $x$ (*округление вниз*).

Величину полиномиального множителя рассчитывают с помощью рядов. Известны асимптотические (аналитические) оценки величины полиномиального множителя для разбиваемых чисел в пределе $n\to\infty$ [15, 16, 17, 18]. Так, асимптотическое выражение Егера имеет, сообразно физическому смыслу задачи, вид [17]

$$\kappa(Q_j, L(N), \eta) = \frac{(\eta+1)^{L(N)}}{\sqrt{\pi\cdot L(N)((\eta+1)^2-1)/6}} \exp\left[\frac{-(Q_j-L(N)\eta/2)^2}{L(N)((\eta+1)^2-1)/6}\right], \tag{22}$$

где $Q_j$ — полиада с меткой $j$; $L(N)$ — количество мод; $\eta$ — наибольшее возможное вибронное число в моде.

*Во всех перечисляемых далее результатах статистическая структура композиций определяется для предела разбиваемого числа $n\to\infty$ и описывает произвольно выбранную композицию из набора всех возможных.*

**Наибольший размер части.** Оценка статистики для наибольшего размера части в прямоугольных композициях нам не известна. В композициях без ограничений Хван и Йе [9] дают оценку такого размера

$$r_{\kappa,макс}(n) = \log_2 n - 3/2 + \gamma/\ln 2 + \mathbf{O}(n^{-1}\ln n), \tag{23}$$

где $\gamma = 0{,}577\ldots$ — *постоянная Эйлера*; $\mathbf{O}(\ldots)$ — *O-нотация* [19, §9.2]; малые по порядку квазипериодичные слагаемые с центральным значением 0 [9] не показаны. Оценку этой же величины в виде

$$r_{\kappa,макс}(n) = \mathbf{\Theta}(\ln n) \tag{24},$$

где $\Theta(...)$ − *тета-нотация* [19, §9.2], подтверждают Бендер и Канфильд [20].

Также Бендер и Канфильд дают статистику для композиций, в которых величины переходов (разностей между размерами последовательно расположенных частей) образуют некое множество чисел $\mathfrak{D}$. Если множества положительных и отрицательных величин переходов конечны (что соответствует одному из двух ограничений, налагаемых физическим смыслом нашей задачи в представлении локальных мод), то наибольший размер части асимптотически равен [20]

$$r_{к,макс}(n) = \Theta([\ln n]^{1/2}). \qquad (25)$$

**Количество различных размеров частей.** Количество различных размеров частей для композиций без ограничений асимптотически равно [9, 21]

$$d_к(n) = \log_2 n - 3/2 + \gamma/\ln 2 + \mathbf{O}(n^{-1} \ln n), \qquad (26)$$

и для случая ограниченных сверху размеров частей [20]

$$d_к(n) = \Theta([\ln n]^{1/2}). \qquad (27)$$

**Структура размеров частей.** Сумма всех различных размеров частей в композиции без ограничений асимптотически равна [9]

$$s_к(n) = (1/2)(\log_2 n)^2 + \mathbf{O}(n^{-1}(\ln n)^2),$$

где малые по порядку квазипериодичные слагаемые с центральным значением 0 [9] не показаны.

**Количество частей.** Количество ненулевых частей, составляющих композицию, для композиций без ограничений [9]

$$m_к(n) = \Theta((n+1)/2). \qquad (28)$$

Для композиций числа $n$ с ограничениями на размеры частей количество ненулевых частей распределяется нормально − существуют ожидание $\mu$ и квадратичное уклонение $\sigma$, зависящие только от рода ограничений, налагаемых на размер части, то есть от состава множества $\mathfrak{D}$, и такие, что [20]

$$\mathrm{P}(m_к = x) = \frac{1}{\sqrt{2\pi n \sigma^2}} \Big( \exp[-(x-\mu n)^2/(2n\sigma^2)] + \mathbf{o}(1) \Big), \qquad (29)$$

где $\mathrm{P}(\mathrm{X})$ − вероятность события X; $x$ − вероятное количество частей.
**Зависимость «размер части − количество частей».** Статистическая

зависимость количества частей от их размера в композициях без ограничений [20, §9]

$$q_к(n,l) = \Theta(n \cdot 2^{-(l+1)}), \qquad (30)$$

где $l$ — размер части. В композициях с ограничениями на размер части характер такой зависимости значительно усложняется, но сохраняется пропорция [20, §3,9]

$$q_к(n,l) = \Theta(Q_l n \rho^{l-1}), \qquad (31)$$

где $Q_l$ — множитель, рассчитываемый при зафиксированном $l$ в условиях ограничений, наложенных на композицию; $\rho$ — параметр, зависящий от рода ограничений: $\rho = \rho(\mathcal{D})$, $\rho < 1$; например, для неограниченных композиций с нулевыми частями $\rho = 0{,}5$; без нулевых частей $\rho = 0{,}57$ [20, §10]. Численный эксперимент Бендера и Канфильда [20, §10] даёт для некоторых типов ограниченных композиций величины $Q_l < 1$, уменьшающиеся по мере нарастания $l$ в (31).

**Выводы.** Согласно выводам комбинаторной статистики, «почти все» полиадные блоки в матрице оператора Гамильтона, исключая интервалы самых низких и самых высоких полиадных чисел объекта, будут иметь большую размерность и «почти одинаковый» состав вибронных чисел (без учёта порядка). Наибольшее вибронное число в пределах блока «почти всегда» окажется степенной функцией от логарифма полиадного числа, и частоты возникновения вибронных чисел в пределах блока будут «почти всегда» распределяться соответственно показательной функции от параметра, характерного для комбинаторных ограничений, выводимых, в свою очередь, из физического содержания задачи.

Более точная статистика требует разработки, тем более что некоторые из указанных результатов определены для случая без ограничения на количество частей.

### Комбинаторное усреднение матричных элементов

Согласно качественным выводам предыдущего подраздела допускаем, что можно свести слагаемые матричных элементов гамильтониана к некоему наиболее вероятному виду (в смысле комбинаторной статистики).

В терминах решаемой физической задачи размеры частей композиций в пределах каждого полиадного блока имеют возможные значения $0 \ldots \eta$ и ограничены, в смысле [20], множеством

$$\mathcal{D} = \{-\eta, -(\eta-1), \ldots, -1, 0, 1, \ldots, \eta-1, \eta\}, \qquad (32)$$

и количество частей композиции зафиксировано как $L(N)$.

Тогда зависимость количества частей композиции полиады $p$ от их размера:

$$q_\kappa(p,k,l) = p \cdot Q_0(p) \rho^{l-1},$$

где $0 < \rho < 1$ — неустановленный параметр, характеризующий тип композиции; $Q_0$ — множитель в смысле [20, §9], такой, что

$$Q_0(p) \sum_{l=0}^{r_{\kappa,max}} \rho^{l-1} = 1,$$

из чего следует нормировочное условие

$$Q_0(p) = \left( \frac{\rho^{r_{\kappa,max}+2} - \rho^2}{\rho - 1} \right)^{-1}.$$

Слагаемые матричных элементов (подраздел «Структура матрицы гамильтониана») в таком усреднении приобретают для полиады $p$ следующий вид:

$$\varphi_{c,ag} = A L(N) \frac{Q_0(p)}{r(p)} \sum_{l=1}^{r(p)} \rho^{l-1} \cdot (-4) \cdot (\Upsilon l - l^2), \qquad (33)$$

где $r(p) = r_{\kappa,max}$ — переобозначение для максимального размера части композиции,

$$\varphi_{cd,ag} = A' L_L(N) \frac{Q_0^2(p)}{r_{+1}^2(p)} \sum_{l_i=0}^{r(p)} \sum_{l_j=0}^{r(p)} \rho^{l_i-1} \rho^{l_j-1} \cdot 4 \left( (l_i+l_j)^2 - 2\Upsilon(l_i+l_j) \right), \quad (34)$$

$$\varphi_{md,ag} = \lambda L_L(N) \frac{Q_0^2(p)}{r_{+1}^2(p)} \sum_{l_i=0}^{r(p)} \sum_{l_j=0}^{r(p)} \rho^{l_i-1} \rho^{l_j-1} \cdot \left( \Upsilon(l_i+l_j) - 2 l_i l_j \right). \qquad (35)$$

Далее, элемент внедиагонального слагаемого для одной пары элементов базиса:

$$\phi_{mm,ag,a'g'}(p) = \lambda L_L(N) \times \kappa_{mm}(p) \frac{Q_0^2}{r^2(p)} \sum_{s_i=0}^{r(p)} \sum_{s_j=0}^{r(p)} \Big\{ \rho^{s_i-1} \rho^{s_j-1} \times$$
$$\times \frac{Q_0^2}{r^2(p)} \sum_{t_i=0}^{r(p)} \sum_{t_j=0}^{r(p)} \rho^{t_i-1} \rho^{t_j-1} \Phi_{mm\pm}(t_i, t_j, s_i, s_j) \Big\} \tag{36}$$

где $\kappa_{mm}(p)$, $0 < \kappa_{mm}(p) \leq 1$ — некий множитель, учитывающий (неизвестную) статистику разностей $+1, -1$ в прямоугольных композициях «почти одинакового» состава (в принятом приближении); двойная сумма по $s_i, s_j$ перечисляет сочетания вибронных чисел в парах сцепленных мод элемента базиса, из которого осуществляется переход; двойная сумма по $t_i, t_j$ перечисляет сочетания вибронных чисел в парах сцепленных мод элемента базиса, в который осуществляется переход; $\Phi_{mm\pm}(...)$ — усреднение выражений (15). Аналитическое выражение для полной суммы в (36) недоступно даже в случае самых простых распределений $\rho$, хотя можно предполагать, — без дальнейшего исследования, — что конечная величина будет включать дробные степени $\ln p$.

Матричные элементы гамильтониана в этом приближении:

$$h_{ag,a=g} = \varphi_{c,ag} + \varphi_{cd,ag} + \varphi_{md,ag}, \tag{37}$$

$$h_{ag,a \neq g} = (\kappa(p)-1) \phi_{mm,ag,a'g'}, \tag{38}$$

где $\kappa(p)$ — количество прямоугольных композиций числа $p$.

## Решение

Такое заполнение матрицы гамильтониана разрешает получить приближение для собственных значений в полиаде $p$ как результат частного случая задачи на собственные значения в следующем виде: 1) однократное решение

$$\varepsilon_{poly,s}(p) = \{\overline{\varphi}_c(p) + \overline{\varphi}_{cd}(p) + \overline{\varphi}_{md}(p)\} + (\kappa(p)-1)\overline{\phi}_{mm}(p) \tag{39}$$

с кратностью 1 и 2) кратное решение

$$\varepsilon_{poly,m}(p) = \{\overline{\varphi}_c(p) + \overline{\varphi}_{cd}(p) + \overline{\varphi}_{md}(p)\} - \overline{\phi}_{mm}(p) \tag{40}$$

с кратностью $\kappa(p)-1$, где $\kappa(p)$ — размерность блока для полиады $p$.

Произвольность, с которой приходится задавать некоторые части слагаемых матричных элементов, а также то, что весовые множители инвариантов $A'$ и $\lambda$ в целом неизвестны и должны определяться особо, не разрешает в дан-ный момент далее развить это решение. Для целей нашей задачи важно то, что 1) используемое комбинаторное усреднение вводит в собственные значения поправки, включающие натуральный логарифм полиадного числа в дробных и целых степенях, и что 2) в выражения для собственных значений явно входят множитель $N$, т. е., количество структурных единиц в объекте, а также функции от него — в составе количественных характеристик структуры $L(N)$, $L_L(N)$, и что 3) множитель $N$ входит в выражение косвенно, через верхние пределы суммирования комбинаторных характеристик.

Таким образом, обобщённый вид выражения для собственного значения энергии объекта, соответствующего полиадному числу $p$,:

$$\varepsilon = N \Phi_N(R(N)) \Phi_p(p, N, A, A', \lambda), \qquad (41)$$

и после приведения к одной структурной единице (атому):

$$\varepsilon_a = \Phi_N(R(N)) \Phi_p(p, N, A, A', \lambda), \qquad (42)$$

где, между прочим, выражается нарушение условия аддитивности [2, §24].

Можно попытаться выразить собственное значение только через известные физические параметры и через поправочные слагаемые, а именно,

$$\varepsilon_a(N, A, Y, p) = A R(N) \big( \Phi_p(p, Y) + \Phi_{pp}(p, Y) \big), \qquad (43)$$

где $\Phi_p(...)$ — функция от дополнительного квантового числа $Y$ и от полиадного числа $p$, являющаяся результатом раскрытия суммы в выражении (33):

$$\Phi_p(p, Y) = \frac{(4 A L(N))}{r(p) \rho^3 (\rho^{r(p)} - 1)} \Bigg( \frac{Y(\rho^{r(p)+1}(r(p)\rho - r(p) - 1) + \rho)}{(1-\rho)} - \\ - \frac{\rho(\rho+1) - \rho^{r(p)+1}\big((r(p)\rho)^2 - (2r(p)^2 + 2r(p) - 1)\rho + (r(p)+1)^2\big)}{(1-\rho)^2} \Bigg). \quad (44)$$

Слагаемое $\Phi_{pp}(...)$ — поправочная функция от тех же параметров, имеющая примерный вид

$$\Phi_{pp} = R(N) \big( \pi_2 p^2 + \pi_1 p + l_2 (\ln p)^2 + l_1 \ln p + c_0 \big), \qquad (45)$$

где $\pi_k, l_k, c_0$ — некие поправочные множители; домножение на размерный множитель $R(N)$ производится согласно общему виду размерной зависимости количества взаимодействий.

Выражения для прочих диагональных слагаемых получаются аналогичным образом и здесь не приводятся из-за громоздкости; выражения для вне-диагональных слагаемых аналитическим образом получить нельзя, но их суммирующая часть может быть оценена, для реальных значений пределов физических параметров локальных мод, как $\mathbf{O}(\Upsilon^{3/2} r(p)^{3/2})$.

Физический смысл поправочной функции $\Phi_{pp}(\ldots)$ — некое изменение, налагаемое на энергетику совокупности невзаимодействующих локальных мод при нахождении (бытии) их в составе объекта. При выражении этой зависимости в виде (45) все поправочные множители соотносятся с силой потенциала одиночной моды $A$.

Подгонка термостатистически усредняемой энергии, рассчитываемой на основе выражения (44), может производиться, например, к дебаевской модели [2, §66], имеющей предельный случай $3 k_B T N$ в макроскопическом состоянии $R(N) \to \mathcal{R}$ при достаточно высоких температурах.

Вообще говоря, разработка представления, выраженного в (44), затруднена, во-первых, тем, что используемое (оценочное) комбинаторное усреднение само разработано недостаточно, во-вторых, использование (43) требует расчёта сумм с большим количеством слагаемых.

Выражение (44) имеет структуру, схожую со структурой выражения для одиночной локальной моды (4). Можно попытаться построить *одномодовое выражение* на основе (4), добавляя в него поправочную функцию, подобно тому, как это сделано в (43, 44):

$$\varepsilon_{1q}(N, A, \Upsilon, n) = A R(N) \big( [-4 \Upsilon n + 4 n^2] + \Phi_{nn}(n, N) \big), \qquad (46)$$

где $n$ — квантовое (вибронное) число в локальной моде; $\Phi_{nn}(\ldots)$ — поправочная функция,

$$\begin{aligned}\Phi_{nn}(n, N) = &q_1 n + q_2 n^2 + q_0 + \\ &+ R(N) \big( q_{l1} \ln(e + n) + q_{l2} (\ln(e + n))^2 \big)\end{aligned}, \qquad (47)$$

где $q_k$ — поправочные множители; *эйлерова постоянная $e$* прибавлена к аргументам логарифмов для смещения нуля поправочных слагаемых.

Подгонка термостатистически усредняемой энергии, рассчитываемой на основе выражения (46), производится, например, к дебаевской модели, как и в предыдущем случае. Такая подгонка делалась для ряда веществ, для которых известны параметры потенциальной функции Морза; получена быстрая и хорошая сходимость.

Окончательно, выражение для канонической статистической суммы в неаддитивном случае для вибронной модели, построенное на основании предлагаемого *комбинаторного одномодового приближения*, имеет вид

$$Z_{1q}(N,T,...) = \sum_{n=1}^{\eta} \exp[-\beta\varepsilon_{1q}(N,A,\Upsilon,n)], \qquad (48)$$

и выражение для внутренней энергии, приходящейся на структурную единицу,

$$u_{1q}(N,T,...) = R(N) \times \left\{\sum_{n=1}^{\eta} \exp[-\beta\varepsilon_{1q}]\right\}^{-1} \times \sum_{n=1}^{\eta} \varepsilon_{1q} \cdot \exp[-\beta\varepsilon_{1q}]. \quad (49)$$

Итоговые выражения, как видно, содержат размерную зависимость, формально выраженную здесь как зависимость от количества структурных единиц $N$. Суммирование по всем возможным полиадам фактически заменено здесь на суммирование «ступеньками», что потенциально огрубляет результат.

Выражение (48) может непосредственно использоваться в выражениях для статистических сумм объекта в нанотермодинамике Хилла (1, 2), т. е.,

$$\Xi_{NPT} = \sum_{\nu}\left\{\exp(-\beta p v_{\nu}) Z_{1q}^{L(N)}\right\}; \quad \Xi_{NVT} = Z_{1q}^{L(N)}. \qquad (50)$$

## Выводы

Можно считать, что в каком-то смысле полученный результат несоразмерен изначально поставленной задаче — комбинаторная статистика, использованная в решении, (пока) разработана в математике недостаточно; соответственно, решение с помощью поправочных слагаемых, хотя и не лишено физического смысла, но слишком «феноменологично», в том смысле, в котором о таких вещах говорит Пайерлс [22]. Всё же, практическое использование полученного результата (излагаемое в других статьях) даёт основания полагать его достаточно корректным и полезным.

## Список использованных источников